\documentclass[useAMS,usenatbib]{mnras}
\usepackage{graphicx,natbib,amssymb,amsmath,stmaryrd,makecell}
\usepackage{txfonts}
\usepackage{xcolor}
\usepackage{hyperref}
\usepackage{xspace}
\usepackage{caption}

\newcommand{\beq}{\begin{equation}}   %

\newcommand{\eeq}{\end{equation}}   %

\newcommand{\beqa}{\begin{eqnarray}}   %

\newcommand{\eeqa}{\end{eqnarray}}   %

\newcommand{\beal}{\begin{align}}

\newcommand{\enal}{\end{align}}

\newcommand{\bspl}{\begin{split}}

\newcommand{\espl}{\end{split}}

\newcommand{\bsub}{\begin{subequations}}

\newcommand{\esub}{\end{subequations}}

\newcommand{\bmulti}{\begin{multline}}   %

\newcommand{\beqm}{\begin{mathletters}}   %

\newcommand{\eeqm}{\end{mathletters}}   %

\newcommand{\changeJ}[1]{{\textcolor{magenta}{#1}}}

\newcommand{\COBEF}{{\it COBE/FIRAS}\xspace}

\author[Acharya, Cyr and Chluba]
{Sandeep Kumar Acharya$^1$\thanks{E-mail:sandeep.acharya@manchester.ac.uk}, Bryce Cyr$^1$ and Jens Chluba$^1$
\\
$^1$Jodrell Bank Centre for Astrophysics, School of Physics and Astronomy, The University of Manchester, Manchester M13 9PL, U.K.
}
\date{\vspace{-0mm}Accepted XXX. Received YYY; in original form ZZZ}

\title[Broad photon spectrum injections]
{Constraining broad photon spectrum injections from exotic and astrophysical sources }

\begin{document}

\maketitle

\begin{abstract}
We study the evolution of photon injections with a power-law type spectrum inserted at various epochs of the universe, and obtain constraints on their parameter space from multiple different cosmological probes. Our work is motivated by the realistic possibility of having extended photon spectra from astrophysical and exotic sources. Going beyond a $\delta$-function like approximation, the physics becomes richer and the constraining power of cosmological probes starts to depend on the photon injection history in a complex way. As a toy model, we first consider a decaying particle scenario, and then generalize to a more model independent power law type injection in redshift. Different combinations of our parameters can be mapped to a wide variety of realistic astrophysical and exotic sources, providing useful benchmarks for study in future work.
\end{abstract}


\section{Introduction}

Cosmological signatures of photon injection from exotic sources such as dark matter decay and annihilation have been a longstanding subject of interest \citep{Sarkar1984,Hu1993b,ASS1999,Chen2004, Padmanabhan2005}, due to their ability to act as a complementary probe to more traditional collider searches. Precise cosmological observations, as well as the ability to probe broad regions of parameter space for various particle physics models, has stimulated further interest in recent years \citep{KK2008,Slatyer2009,Chluba2010a, Chluba2015GreensII,poulin2015,poulin2016,AK2019,Bolliet2020PI, Liu2023II,CFLM2023}. In these works, it is typically assumed that the dark matter decays or annihilates to photons as a $\delta$ function in energy, after taking kinematic constraints into account.
In reality, however, the process might be more complicated and one can expect broad distributions of injected photons, e.g., from the reprocessing in the plasma. For these scenarios, the effects on various cosmological observables can be significantly modified, and the constraints should be recomputed appropriately.

At early times, the Cosmic Microwave Background (CMB) photons and electrons are usually in thermal equilibrium, and injection of non-thermal photons will disturb that equilibrium. In an expanding universe, the re-thermalization process will be incomplete depending upon the epoch of photon injection \citep{Zeldovich1969,Illarionov1975b,Chluba2011therm,Khatri2012mix, Chluba2015GreensII}. The deviation of the CMB spectrum from a Planckian is an observable which can be constrained using the {\it COBE/FIRAS} \citep{Fixsen1996} experimental data. Injection of energetic photons ($E \gtrsim 13.6$ eV) around the recombination epoch ($z\simeq 10^3$) and afterwards, can ionize neutral hydrogen, resulting in delayed recombination and a higher freezeout electron fraction. Consequently, CMB photons scatter off of electrons with a higher probability, which reduces the CMB temperature anisotropies while boosting those in polarization. These effects can be constrained with CMB anisotropy data as measured by {\it Planck} \citep{Planck2018params}. 

The 21 cm distortion signal provides a complimentary probe of photon injections at redshifts much lower than that of the recombination epoch.
Additional photon injections on top of the CMB will add to the existing radio background, which in turn will increase the 21 cm absorption depth \citep{F2006,FH2018}. 
Current and future global 21 cm experiments are hunting for this effect, and if the signal has a sufficiently deep absorption trough, a confirmed observation may be on the horizon.

However, to accurately predict the imprint of excess radio backgrounds one has to carefully treat the absorption of soft photons ($E \lesssim 13.6$ eV) by the free-free process. In fact, this absorption leads to extra heating of the gas, which in turn produces a shallower absorption feature, even in the presence of an enhanced radio background. This allows for the hiding of significant radio backgrounds in the 21 cm absorption signature, contrary to the common expectation in the literature \citep{ACC2023}.
Additionally, the radio background can be measured and any excess signal on top of what is expected from standard astrophysical sources may hint at a new source of non-thermal photons. There indeed has been a detection of an unexplained radio excess by the ARCADE-2 \citep{Fixsen2011} experiment at frequencies of $3-90$ GHz. At present, our knowledge of expected astrophysical radio sources cannot explain the detected excess \citep{GTZBS2008}, though significant uncertainties still exist \citep{COGSHHL2023}. One physical solution to this radio excess that does not violate existing constraints could be related to the photon production from superconducting cosmic string networks \citep{Cyr2023CSS}, a scenario that could be further tested with upcoming CMB spectrometers such as TMS \citep{Jose2020TMS} and BISOU \citep{BISOU}. A broad discussion of various interpretations and models can be found in \citet{Singal2018, Singal2022}.

In this paper we wish to extend the analysis of photon injection scenarios to include broad photon distributions. Specifically, we consider the injected photons to be power-law distributed in frequency and derive constraints on the parameter space using CMB (anisotropy and distortion), 21 cm and radio background data. As a toy model, we consider decaying particle scenarios to fix the time-dependence of the injection. However, we also consider a more model-independent scenario where the photons are injected according to a power-law in redshift, which can map to a broader class of astrophysically-motivated scenarios. For a broad photon spectrum, one expects a richer and more complicated impact on the observables, in particular during the post-recombination era. We show how these constraints depend on the shape of the photon spectrum by studying two specific cases with synchrotron and free-free type injections. Such scenarios offer plausible models for the soft photon emission from high energy particle cascades as well synchrotron emission from accreting black holes. However, we should remind the reader that for a generic photon spectrum injection, one needs to compute the constraints on a case-by-case basis. In this work, we have used the code {\tt CosmoTherm} \citep{Chluba2011therm} to derive our constraints. 

A short description of this paper is as follows. We give a brief overview of the cosmological observables that we use in Sec. \ref{sec:observables}. In Sec. \ref{sec:model}, we describe our model for photon injection and follow it up with the discussion of constraints on parameter space in Sec. \ref{sec:decay_constraints}. 
We discuss constraints for the power law (in redshift) injection case in Sec. \ref{sec:plaw_constraints} and end with discussions in Sec. \ref{sec:conclusion}.

\section{Cosmological observables}
\label{sec:observables}

We briefly review the cosmological observables that we use in this work to constrain the parameters of our model. Interested readers can find more details in \cite{CCA2023,ACC2023}. In particular, we use a similar likelihood setup to search for regions that are in tension with existing observations.

\subsection{CMB spectral distortions}
Injection of photons or pure energy release to the baryon-fluid plasma in the pre-recombination epoch can distort the CMB as a result of the non-equilibrium between CMB photons and the electrons. Depending upon the redshift at which this energy is injected, a $y$ or $\mu$-distortion \citep{Zeldovich1969,Illarionov1975b,Chluba2011therm} can be created. While $y$-distortions are produced in the low optical depth regime ($z\lesssim 10^4$), these distortions relax to the $\mu$-type for $z\approx 2\times 10^5$. 
We compute the spectral distortion solutions using the numerical code {\tt CosmoTherm}, which follows all the important thermalization processes \citep{Chluba2011therm}. 

The current upper limit on the amplitude of the $y$ and $\mu$ parameters are $|y|\lesssim 1.5\times 10^{-5}$ and $|\mu|\lesssim 9\times 10^{-5}$, at $2\sigma$ \citep{Fixsen1996}, with discussions of improved constraints in \citet{tris2, Bianchini2022}.
However, the limits depend upon the specific shape of the induced spectral distortion since the {\it COBE/FIRAS} data places constraints on CMB distortions within $60-600$ GHz. In principle, soft photon injections at $\nu\lesssim 60$ GHz in the post-recombination era can have drastically different constraints. In this work, we compute the spectral shapes using {\tt CosmoTherm} for each parameter combination and use the {\it COBE/FIRAS} data to place constraints on photon injection cases. The reader can find the details of this procedure in \cite{CCA2023}.

\subsection{CMB anisotropy}
\label{subsec:cmb_anisotropy}
Direct injection of sufficiently energetic electrons or photons ($\gtrsim 13.6$ eV) during and post-recombination, as well as heating due to pure energy release, will result in a higher abundance of free electrons at $z\lesssim 10^3$. The increased probability for CMB photons to scatter off of free electrons damps CMB temperature anisotropies, while boosting the polarization signals \citep{ASS1999,Chen2004}. Precise observation of these anisotropies from CMB experiments are capable of \citep{WMAP7yrPower,Planck2018params} placing strong constraints on these types of energy injection cases \citep{Galli2009,Slatyer2009,Huetsi2011,AK2019}. 

The recombination history can be computed accurately following previous works \citep{Zeldovich68,Peebles68,Seager2000,Chluba2006, Sunyaev2009,chluba2010b,Yacine2010c}. In this paper, we compute the recombination history, including our toy model for photon injections, using the {\tt Recfast++} \citep{chluba2010b} module within {\tt CosmoTherm} \citep{Chluba2011therm}. This allows us to propagate the effects of extra heating and ionization caused by the additional non-thermal photons.

To compute CMB anisotropy constraints, we use the direct projection method developed in \cite{PCA2020} which is based on a principal component analysis (PCA) of recombination history perturbations first studied in \citet{Farhang2011, Farhang2013}. Using {\tt CosmoTherm}, we compute the fractional change to the standard ionization history of the universe, 
for a given parameter combination within our model. We then compute the first three principal component coefficients by projecting the change to the ionization history onto the eigenmodes of the recombination perturbations. 
Using the covariance matrix of the $\mu_i$ obtained in \cite{PCA2020} as well as updated data from {\it Planck} 2018 \citep{Planck2018params}, we then compute the likelihood of the model assuming Gaussian statistics. With this method, we can obtain CMB anisotropy constraints in a way which can efficiently mimic the results one would expect from a full MCMC analysis.

\subsection{Global 21 cm signatures}
The differential brightness temperature ($\Delta T_{\rm b}$) is the main observable in global $21$ cm cosmology. It is a measure of the contrast between the background radiation temperature at the $21$ cm line ($T_{\rm R}$) and the spin temperature of the gas ($T_{\rm s}$) at a redshift $z$, usually given by \citep[see e.g.][]{F2006}

\begin{equation}
\label{eq:DTb}
    \Delta T_\text{b} = \dfrac{\left(1-{\rm e}^{-\tau_{21}}\right)}{1+z}\left(T_\text{s}-T_\text{R} \right),
\end{equation}
where $\tau_{21}$ is the 21 cm optical depth. Typically, one assumes that $T_{\rm R}=T_{\rm CMB}$, i.e. that the CMB is the only source of radio photons. However, when additional radio photons from exotic sources such as decaying or annihilating dark matter (or even from early time astrophysical sources) are present, the radiation temperature is modified, $T_{\rm R}=T_{\rm CMB}+\Delta T$. In the Rayleigh-Jeans regime, we can write,
\begin{equation}
    \frac{T_{\rm CMB}+\Delta T}{T_{\rm CMB}}\Bigg|=\frac{I_{\rm CMB}+\Delta I}{I_{\rm CMB}}\Bigg|_{1.4 {\rm GHz}},
\end{equation}
where $\Delta I/I_{\rm CMB}|_{1.4 {\rm GHz}}$ is the CMB spectral distortion in the rest frame frequency of 1.4 GHz at $z$ due to the extra photon source term. 

Near the time of cosmic dawn, the spin temperature is largely driven towards the temperature of the ambient hydrogen gas. Previous works neglected the free-free absorption of low energy photons by the gas. This absorption (which we call ``soft photon heating"), actually can lead to a sharp rise in the matter temperature and, subsequently, in the spin temperature for sufficiently steep radio backgrounds. For a detailed discussion on the modelling of the 21 cm signal, and soft photon heating, the reader is referred to \cite{ADC2022,ACC2023}.

For this work, we use the EDGES measurement \citep{Edges2018} as a figure of merit to constrain the various parameters in our model. EDGES claimed a detection of the 21 cm absorption feature with a $\Delta T_{\rm b}\simeq-500$ mK originating from $z\approx 18$ and a 1$\sigma$ error of $200$ mK. 
In this paper, we demand that $-500~{\rm mK}\lesssim \Delta T_{\rm b} \lesssim 0~{\rm mK}$ at $z\approx 18$ for the parameter combinations in our model. Outside of this regime we use a Gaussian likelihood with an error of $200$ mK to quantify the tension with this data. 
We stress that the EDGES detection remains a hot debate in current literature \citep{HKMP2018}. In fact, a recent experiment \citep{Saras2022} has failed to reproduce this detection, highlighting the need for additional independent measurements and analyses.

\vspace{-3mm}
\subsection{The radio synchrotron background (RSB) excess}
The ARCADE-2 experiment \citep{Fixsen2011} has detected a radio background excess (on top of the radio photons expected from the CMB) in the frequency range of $3-90$ GHz.
%
%
These RSB data points, alongside some previous results which were analyzed and compiled in their Table 4, are well fit by a power law with spectral index 2.6 and temperature $T\simeq 24$~K at 310 MHz. In \cite{DT2018}, the authors performed an independent analysis using data points around $\simeq 40-80$ MHz and found the best fit slope to be consistent with ARCADE-2 but with a slightly higher normalization of $\simeq$ $30$K at $310$ MHz. 

It has been demonstrated that even when taking all resolved, extra-galactic sources into account, one is still unable to fully explain this radio excess \citep{GTZBS2008, Tompkins2023}. In this work, we use a fitting function to this minimal extra-galactic background (MEG), given by \citep{GTZBS2008},
\begin{equation}
    T_{\rm bg}(\nu)\simeq 0.23\,{\rm K}\left(\frac{\nu}{\rm GHz}\right)^{-2.7}. 
    \label{eq:extra-galactic}
\end{equation}
Taking this MEG into account, the data analyzed by \citet{Fixsen2011,DT2018} can be fit by a power law \changeJ{as}
\begin{equation}
\label{eq:RSB-fit}
    T_{\rm RSB}(\nu)\simeq 1.230\,{\rm K}\left(\frac{\nu}{\rm GHz}\right)^{-2.555}. 
\end{equation}
We use this model as a reference for our likelihood evaluation. In practice, we add the spectral distortion from our photon injections (computed by \texttt{CosmoTherm}) to the MEG, and require that the low frequency spectrum not exceed the fit given by Eq.~\eqref{eq:RSB-fit}.

\subsection{Optical depth constraints}
Increased free electron density due to photon injections will lead to a higher optical depth during the reionization epoch which can then be constrained using CMB anisotropy data.

We use the optical depth ($\tau$) measurement of \cite{Planck2018params} with $\tau=0.0544\pm 0.0073$ to constrain our energy injection scenarios. We use a simple parametric model for reionization, the details of which can be found in \cite{ADC2022}. We should note that, ideally, the CMB anisotropy constraints discussed in Sec. \ref{subsec:cmb_anisotropy} should include changes to the reionization history. However, the current PCA setup is only sensitive to changes in the reionization history at $300\lesssim z\lesssim 4\times 10^3$ and does not directly model changes to the this history. Therefore, we include a separate constraint from modifications to $\tau$.

In this work, we assume that energy injections modify the reionization history, or $\tau$, only perturbatively. Since our fiducial reionization model gives a $\tau=0.078$,\footnote{For this we assume the starting redshift of reionization be $z=30$, though changing this does not alter our results.} we compute the differences between the optical depth obtained with and without photon injections. We use a simple Gaussian penalty to constrain departures of our model away from this fiducial value. We have checked that tuning the reionization model parameters to more closely reproduce the measured $\tau$ value does not alter the constraints much.

\section{Model for photon injection}
\label{sec:model}

In this work, we consider energy injection in the form of a broad photon spectrum from a toy model of decaying dark matter. Cosmological constraints for such injection cases will be model dependent, however, we want to showcase the basic physics within the setting of a simplified model. For the case of particle decay, the energy injection rate into the CMB is given by,
\begin{equation}
    \frac{{\rm d}\rho_{\gamma}}{{\rm d}t}=f_{\rm dm}\Gamma\rho_c{\rm e}^{-\Gamma t},
    \label{eq:decay}
\end{equation}
where $f_{\rm dm}$ is the fraction of decaying dark matter, $\rho_c$ is the energy density of cold dark matter and $\Gamma$ is the inverse of decay lifetime. Note that $f_{\rm dm}$ can exceed unity for lifetimes sufficiently short such that most of the dark matter decays before matter radiation equality.

In this work, we consider the decay product to be an extended soft photon spectrum with power law shape and a high frequency cutoff. We assume an injected soft photon spectrum of the form,
\begin{equation}
    \Delta I(x)\propto x^{-\gamma}{\rm e}^{-\frac{x}{x_{\rm inj,0}}},
    \label{eq:soft_spectrum}
\end{equation}
where $x=\frac{E}{{\rm k_B}T_{\rm CMB}(z)}$ is the dimensionless photon frequency (with photon energy $E$), and $x_{\rm inj,0}=\frac{E_{\gamma}}{{\rm k_B}T_{\rm CMB,0}}$ is the high frequency cutoff ($T_{\rm CMB,0}$ being the CMB temperature today). The redshifting of photons is implicitly taken into account with the definition of the dimensionless frequency $x$.
 
In addition, we need not restrict ourselves to scenarios in which all of the decay energy goes into soft photons. If we allow a fraction of the total injected energy to be deposited as heat to the background electrons, the temperature evolution equation must be modified \citep{Chluba2010a},
%
\begin{equation}
    \frac{{\rm d}T_b}{{\rm d}t}=\frac{2}{3{\rm k_B}}\frac{1-\epsilon_{\rm soft}}{N_H(1+f_{\rm He}+x_e)}\frac{{\rm d}\rho_{\gamma}}{{\rm d}t}.
    \label{eq:baryon_temperature}
\end{equation}
Here, $N_H$ is the hydrogen number density, $f_{\rm He}$ is the helium number fraction relative to hydrogen and $x_e$ is the free electron fraction.

The branching fraction of energy in the soft photons with respect to the total injected energy is denoted as $\epsilon_{\rm soft}$. We consider free-free and synchrotron spectra as example cases with $\gamma=0$ and 0.6, respectively. We first study the choice of $\epsilon_{\rm soft}=1$, and generalize later on. The variation in $\epsilon_{\rm soft}$ will allow us to showcase the differences between direct photon injection and pure energy release, which is known to be important in the context of 21 cm cosmology \citep{ACC2023}.

\section{Discussion of constraints}
\label{sec:decay_constraints}

We are interested in finding the regions of parameter space ruled out by cosmological observables, and are not necessarily in determining where the best fitting regions are (see \citet{ACC2023, Cyr2023CSS} for this discussion). Our model consists of four parameters, $\Delta\rho_{\gamma}/\rho_{\gamma}$ (the total fractional energy injection to the CMB), $x_{\rm inj,0}$, $\Gamma$ and $\epsilon_{\rm soft}$. Since it is difficult to simultaneously scan over four parameters, we choose to only scan over $x_{\rm inj,0}$ and $\Gamma$ for some reasonable choices of the other parameters. As an example, choosing $\Delta\rho_{\gamma}/\rho_{\gamma}\lesssim 10^{-4}$ allows us to be broadly consistent with {\it COBE/FIRAS}. Similarly, we demand that the injected photon spectrum is extended up to at least 90 GHz, which is the highest frequency probed by the ARCADE-2 experiment \citep{Fixsen2011}. This translates into us demanding $x_{\rm inj,0}\gtrsim 1$. We choose the upper limit of $x_{\rm inj,0}$ to be $10^5$. Above this, we cross the ionization threshold of hydrogen and the constraints become largely independent of $x_{\rm inj,0}$. The value of $\Gamma$ ranges from $10^{-8} \, \rm{s}^{-1} -10^{-18}$ s$^{-1}$ which covers the timescales of universe from the $\mu$ era to the present day. We also require that for lifetimes comparable to or longer than the age of universe, only a tiny fraction of dark matter decays such that the evolution history of the universe does not change significantly. Using the cosmological observables that we consider in this work, the allowed $\Delta\rho_{\gamma}/\rho_{\gamma}$ turns out to be $\lesssim 10^{-4}$, as we will show below. In the case of $\Delta \rho_{\gamma}/\rho_{\gamma} \simeq 10^{-4}$, we have checked that $f_{\rm dm}\lesssim 10^{-7}-10^{-8}$, implying that most of dark matter is still intact.

\begin{figure*}
\centering 
\includegraphics[width=\columnwidth]{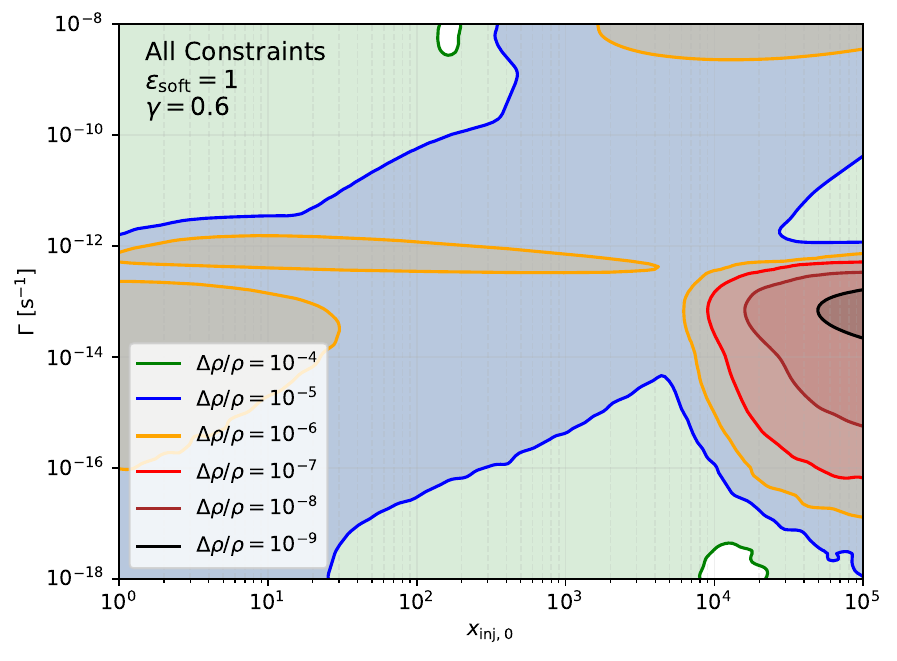}
\hspace{4mm}
\includegraphics[width=\columnwidth]{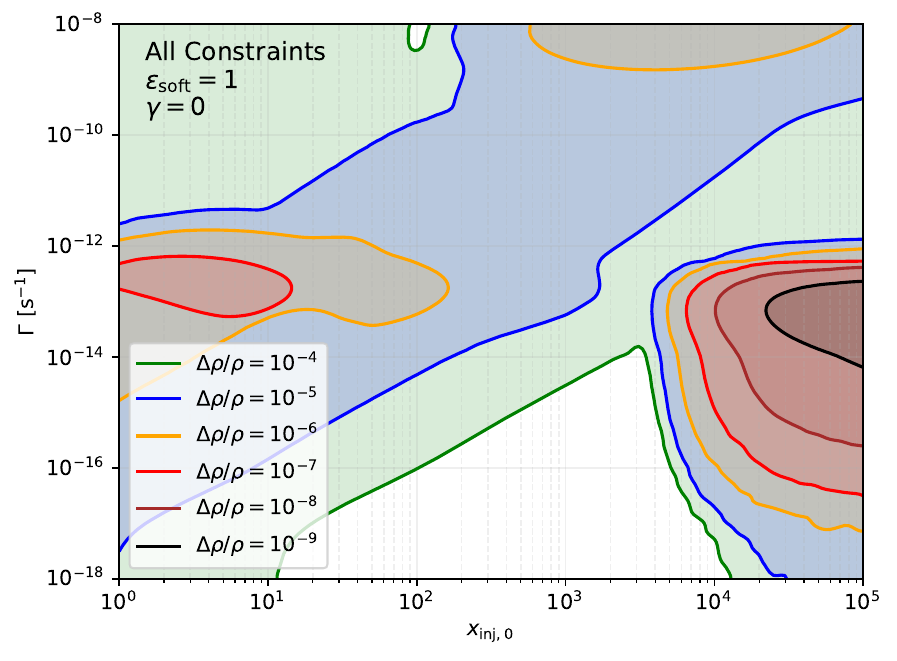}
\\[10mm]
\includegraphics[width=\columnwidth]{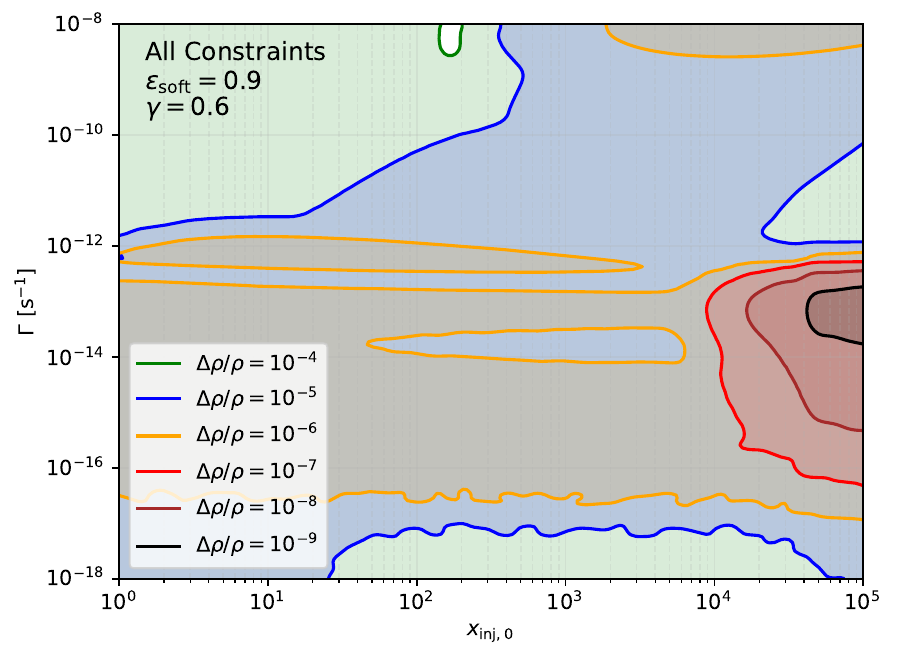}
\hspace{4mm}
\includegraphics[width=\columnwidth]{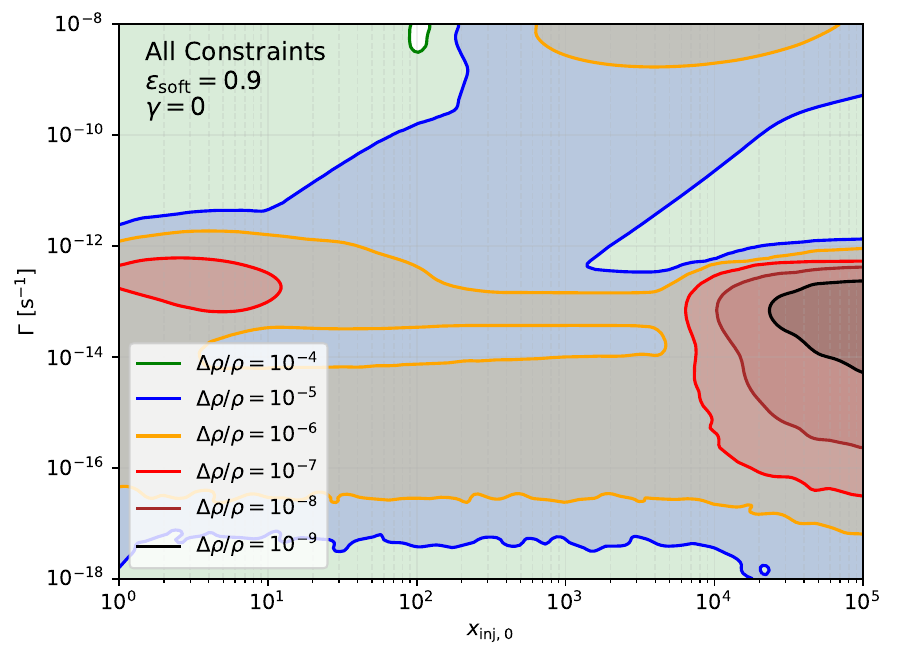}
\\[10mm]
\includegraphics[width=\columnwidth]{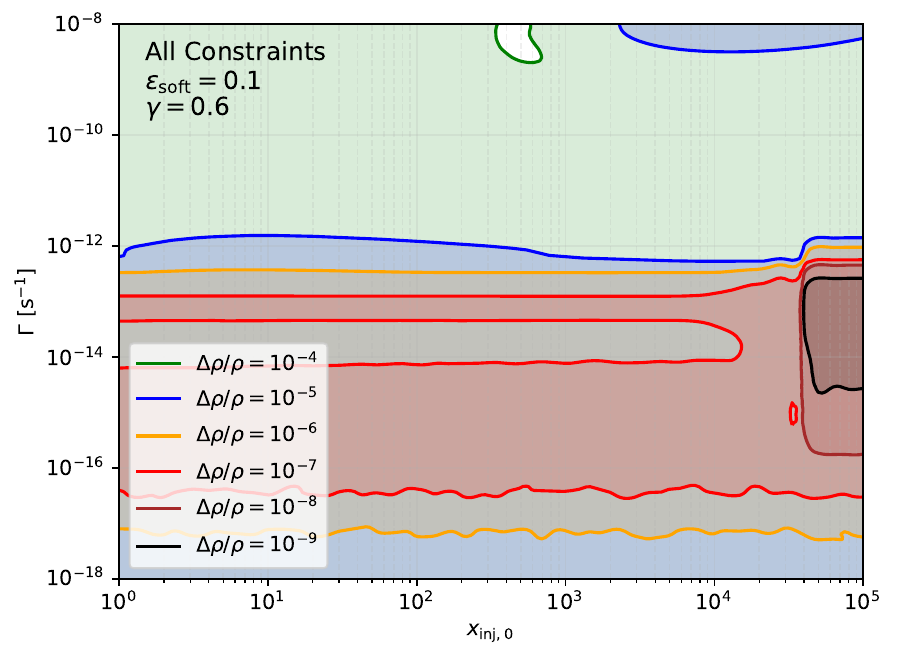}
\hspace{4mm}
\includegraphics[width=\columnwidth]{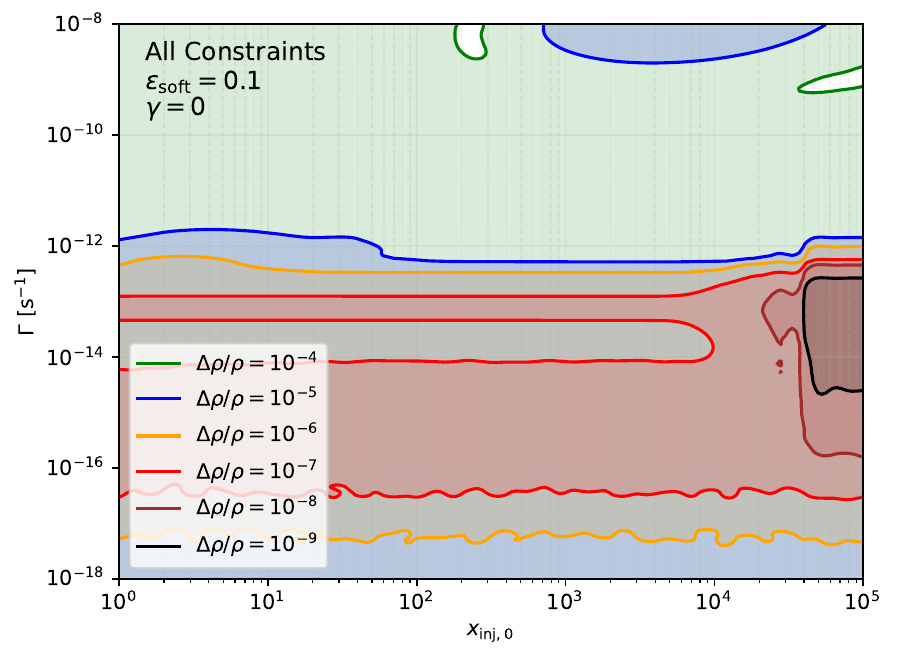}
\caption{A summary of the full constraints on the lifetime $\Gamma$ and high frequency cutoff $x_{\rm inj,0}$ for a variety of benchmarks values of $\Delta \rho_{\gamma}/\rho_{\gamma}$. We use \texttt{CosmoTherm} to compute limits based on CMB anisotropies, global 21 cm signals, extra radio backgrounds, and modifications to the optical depth to reionization ($\tau$). The left panels are for a synchrotron type injection spectrum, and the right for free-free type. A breakdown of the constraints can be found in subsequent figures.}
\label{fig:total_const}
\end{figure*}

\begin{figure*}
\centering 
\includegraphics[width=\columnwidth]{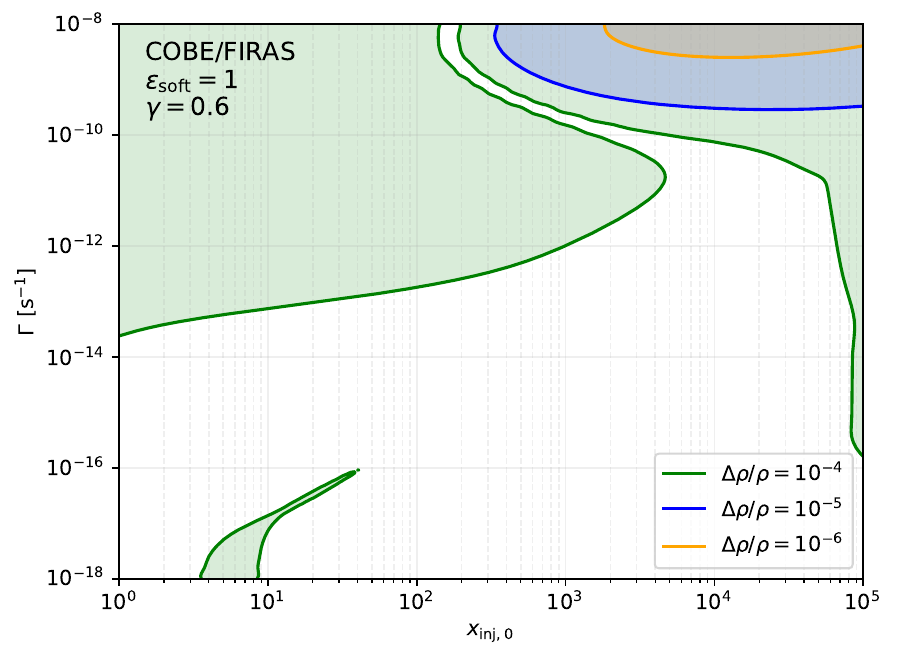}
\hspace{4mm}
\includegraphics[width=\columnwidth]{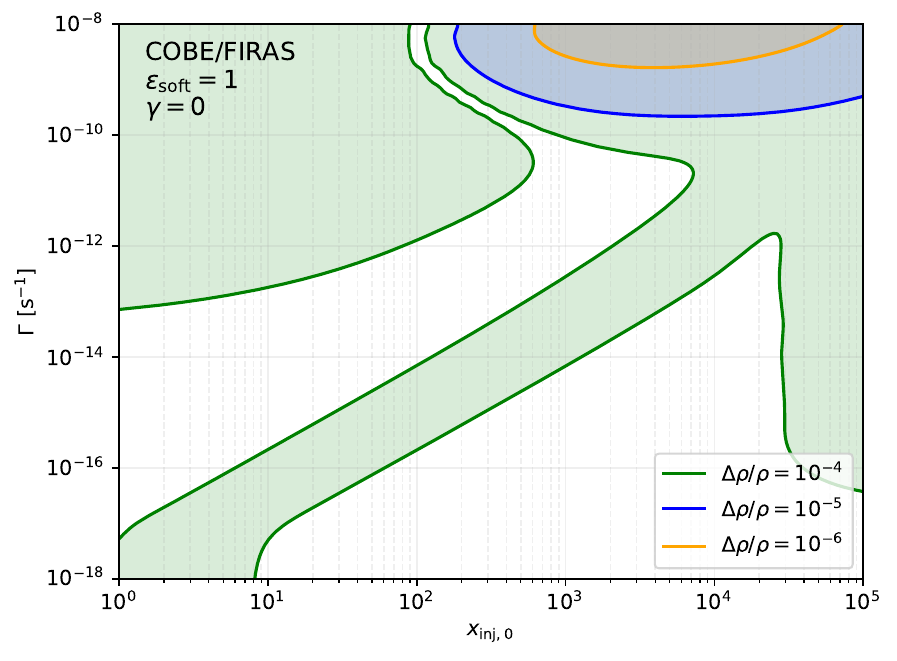}
\\[10mm]
\includegraphics[width=\columnwidth]{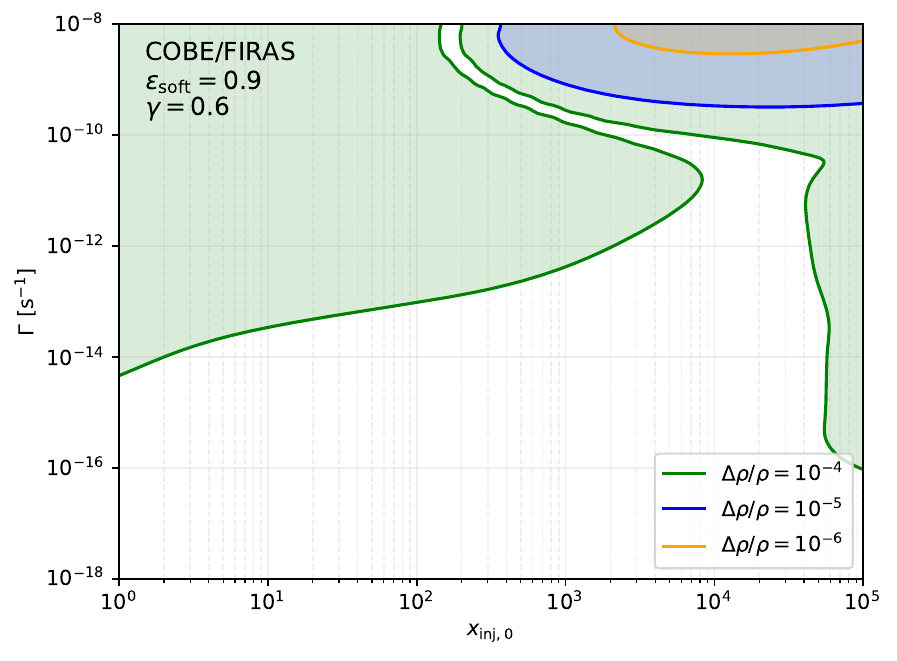}
\hspace{4mm}
\includegraphics[width=\columnwidth]{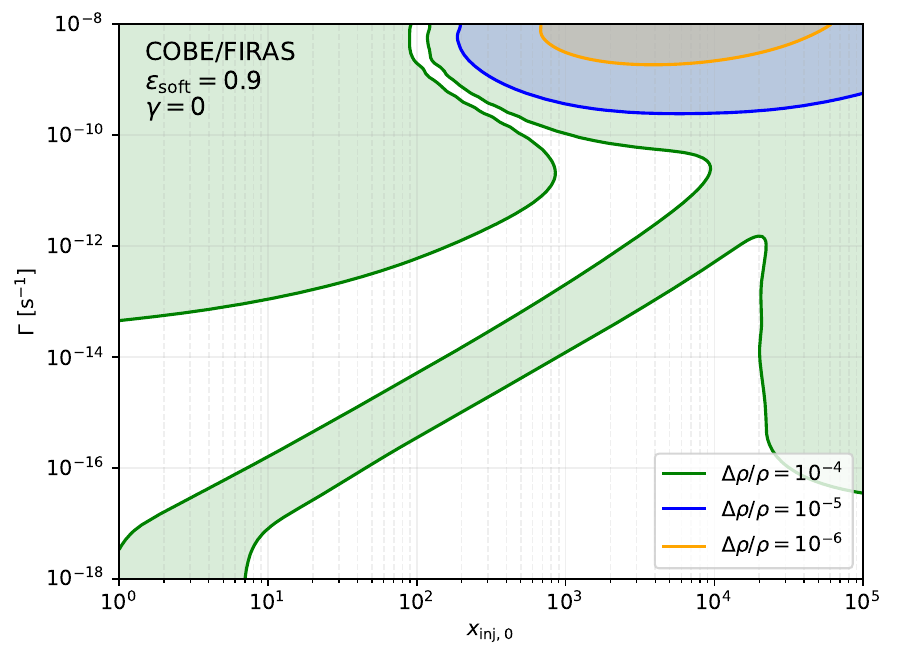}
\\[10mm]
\includegraphics[width=\columnwidth]{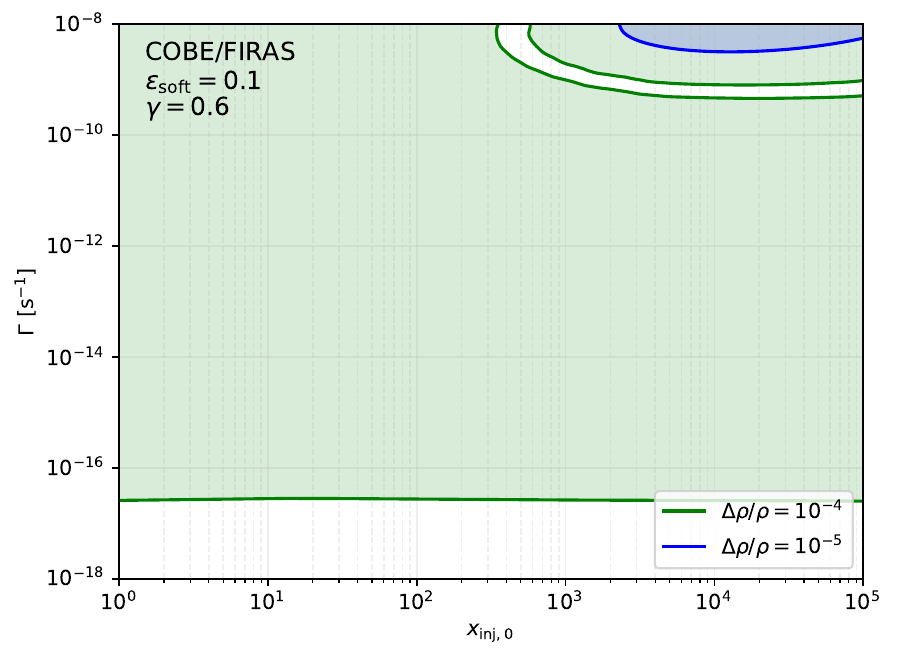}
\hspace{4mm}
\includegraphics[width=\columnwidth]{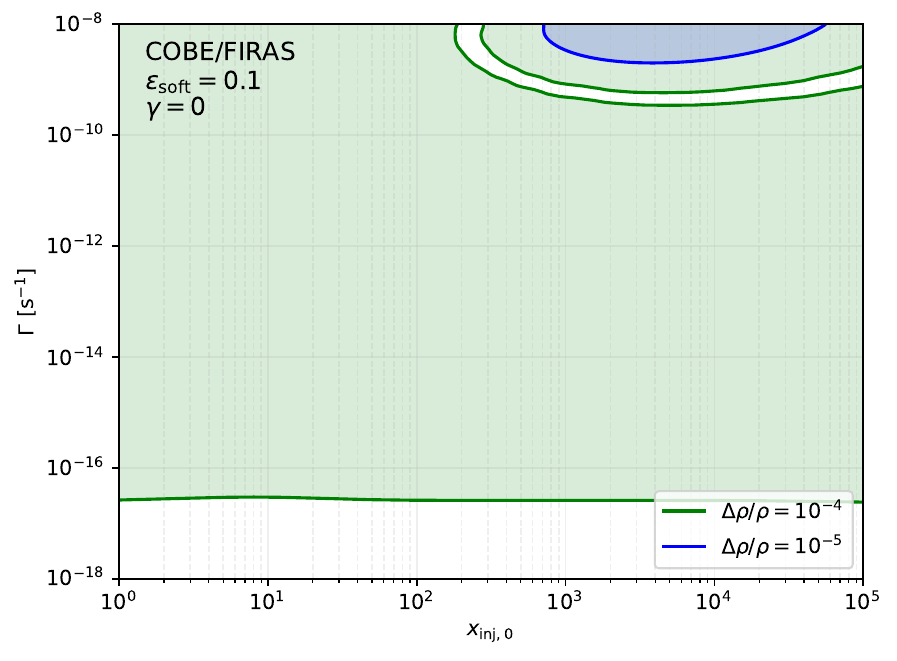}
\caption{Limits from {\it COBE/FIRAS} on synchrotron (left) and free-free (right) type photon injections. The unconstrained strip in the top right of the plots comes from a convenient cancellation of the induced spectral distortion signatures. For small $\epsilon_{\rm soft}$, \COBEF is only capable of constraining pure energy injections (heat) for large values of $\Delta \rho_{\gamma}/\rho_{\gamma}$.}
\label{fig:FIRAS_const}
\end{figure*}

\begin{figure*}
\centering 
\includegraphics[width=\columnwidth]{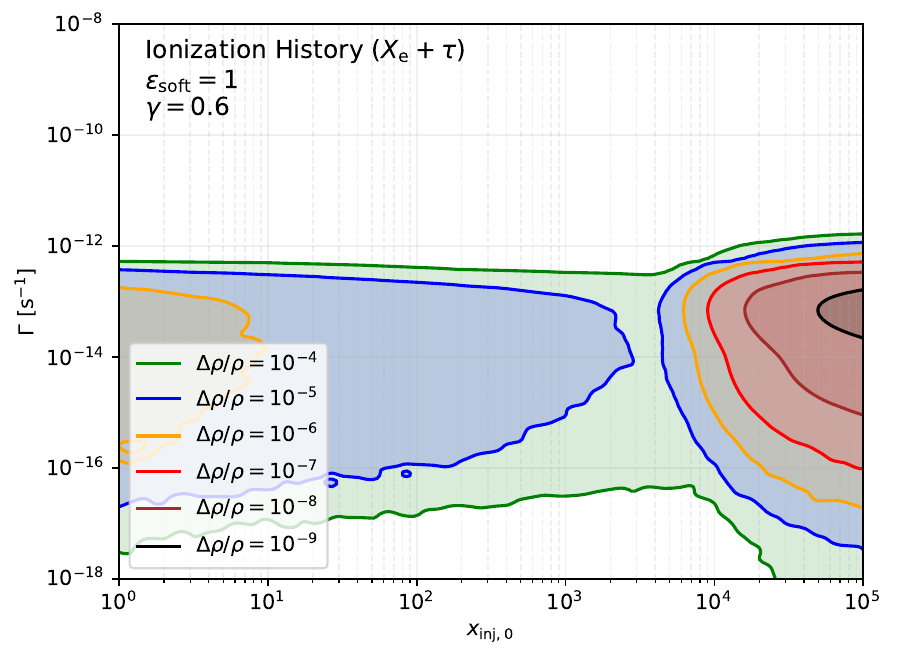}
\hspace{4mm}
\includegraphics[width=\columnwidth]{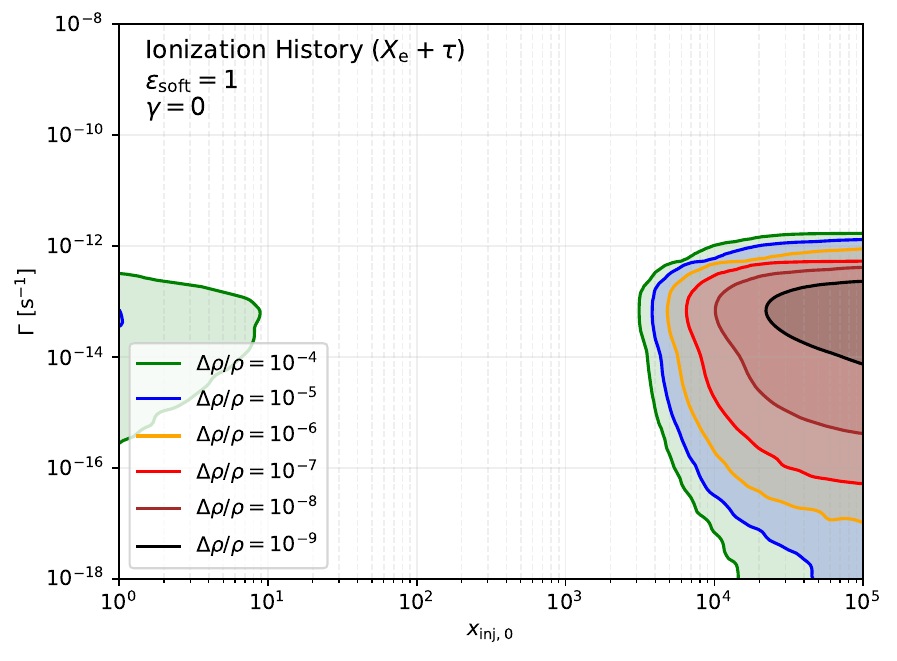}
\\[10mm]
\includegraphics[width=\columnwidth]{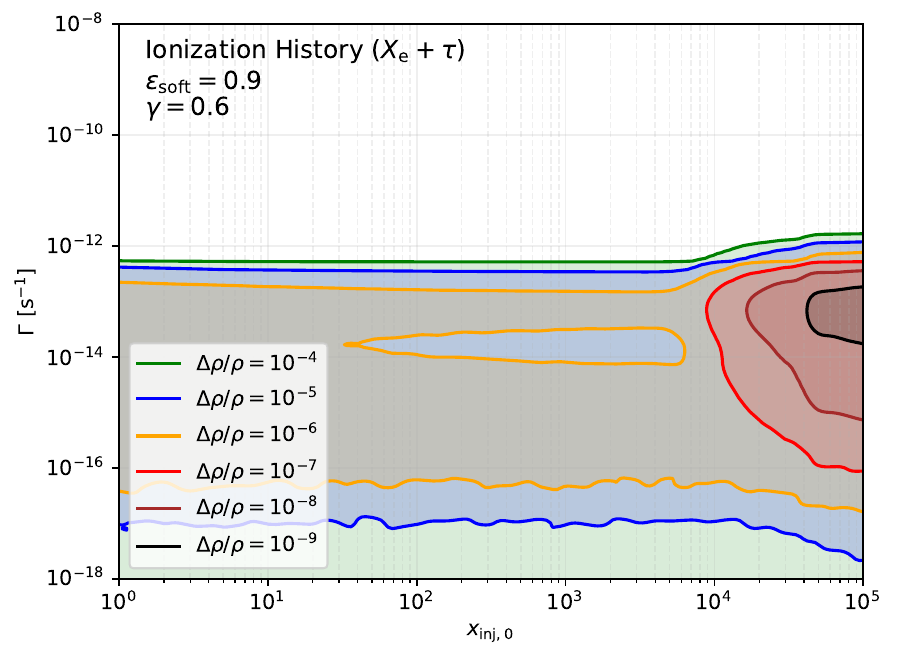}
\hspace{4mm}
\includegraphics[width=\columnwidth]{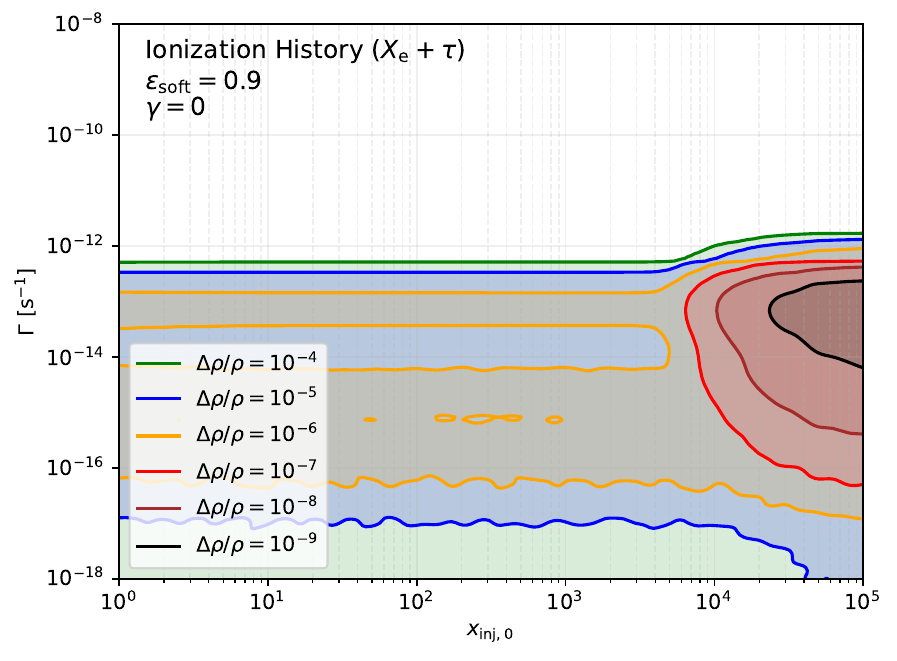}
\\[10mm]
\includegraphics[width=\columnwidth]{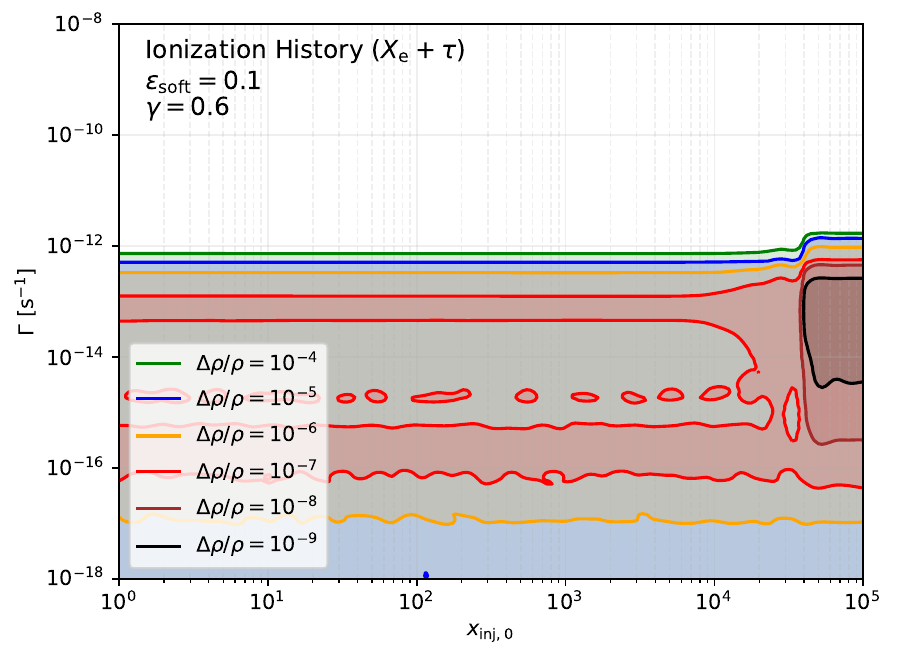}
\hspace{4mm}
\includegraphics[width=\columnwidth]{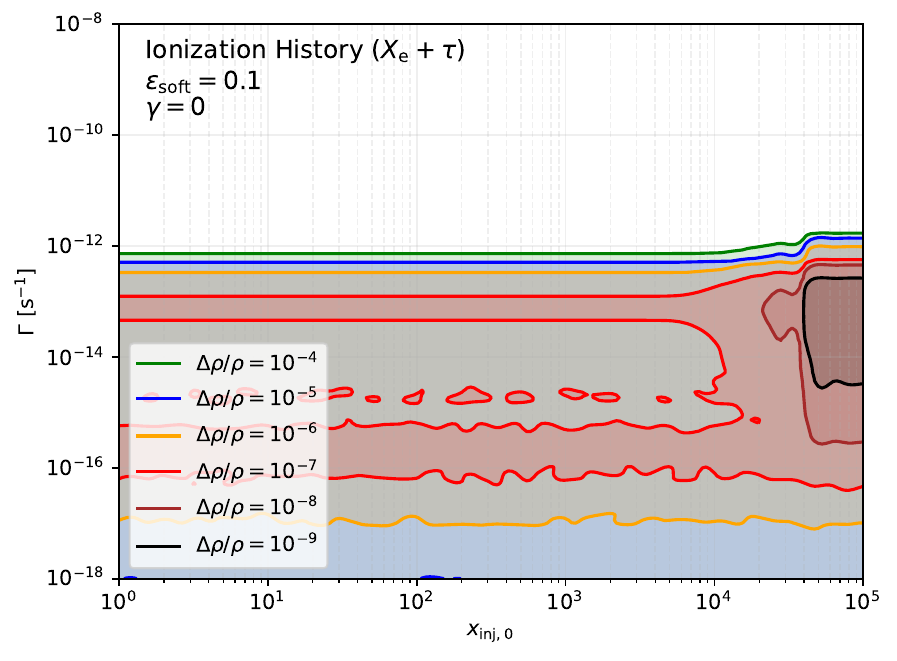}
\caption{Changes in the free electron fraction ($X_{\rm e}$) and the optical depth to reionization ($\tau$) are capable of providing stringent constraints when the photon/energy injections take place near and post-recombination ($\Gamma \lesssim 10^{-13} \, {\rm s}^{-1}$).}
\label{fig:XeTau_const}
\end{figure*}

\begin{figure*}
\centering 
\includegraphics[width=\columnwidth]{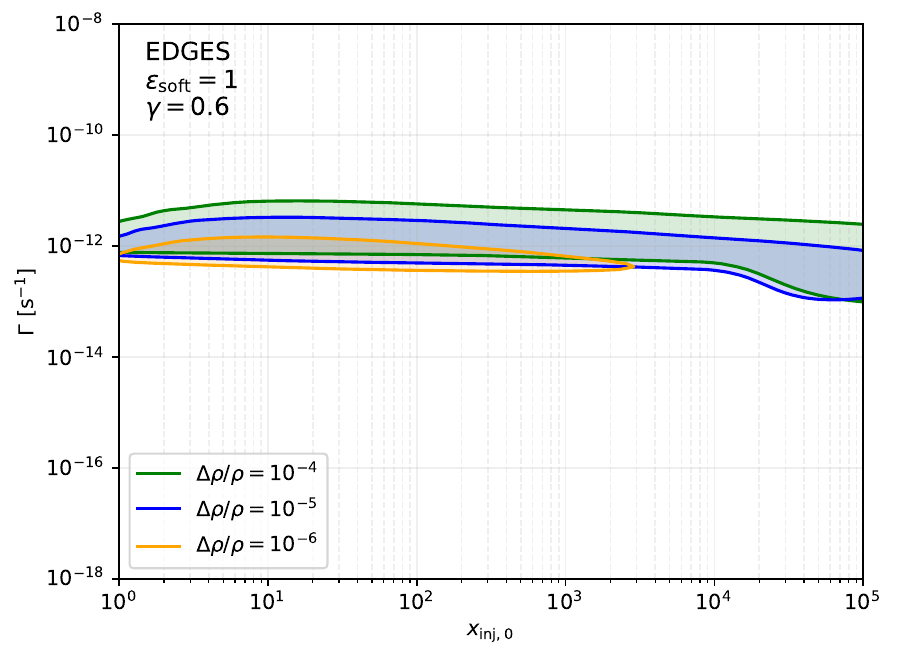}
\hspace{4mm}
\includegraphics[width=\columnwidth]{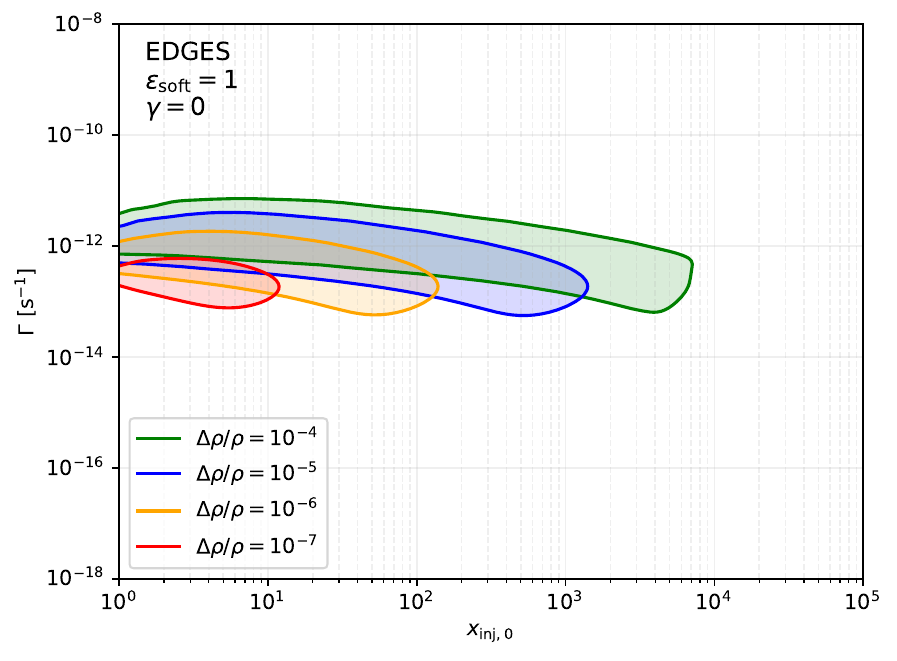}
\\[10mm]
\includegraphics[width=\columnwidth]{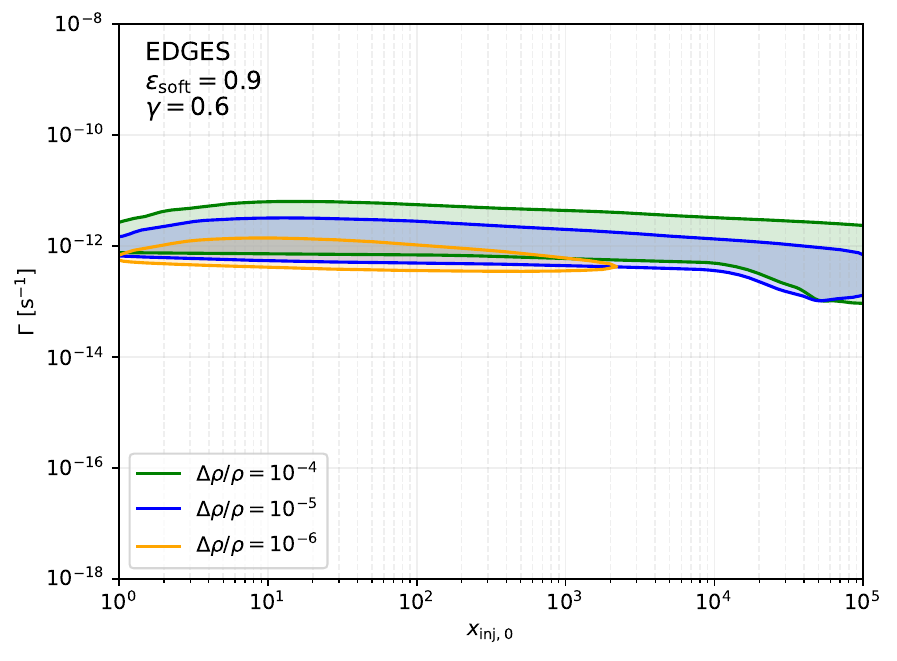}
\hspace{4mm}
\includegraphics[width=\columnwidth]{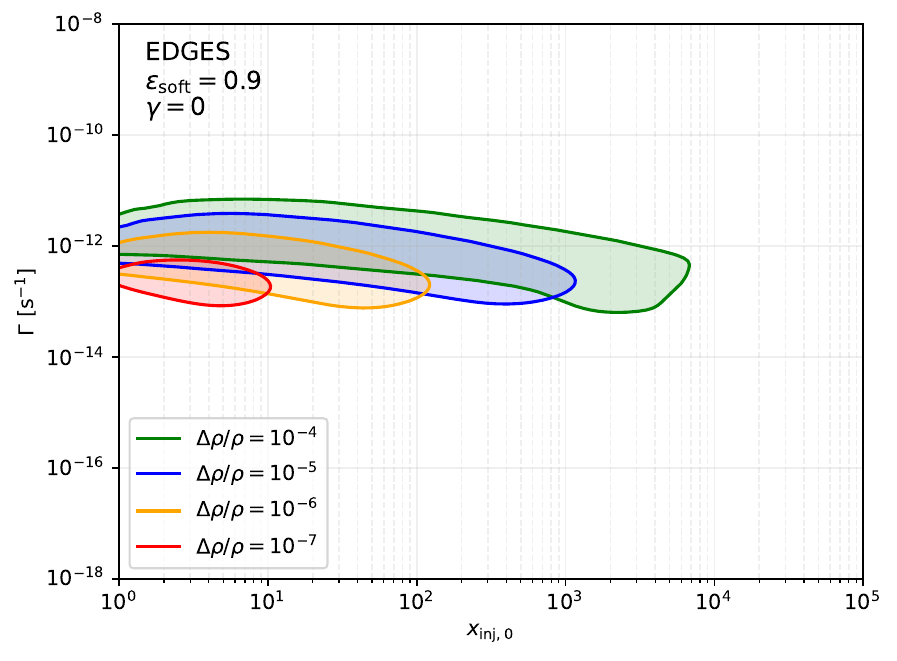}
\\[10mm]
\includegraphics[width=\columnwidth]{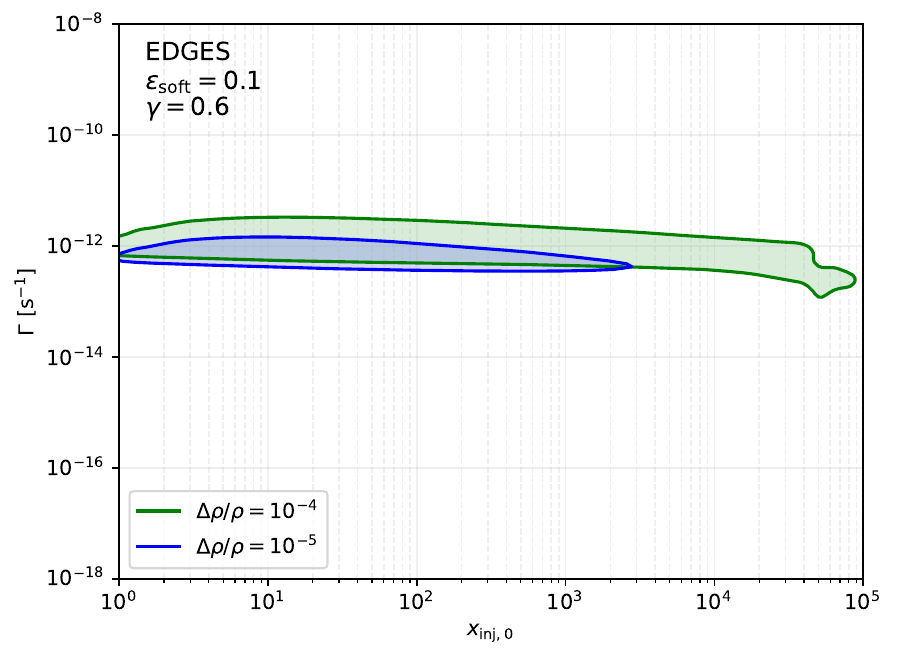}
\hspace{4mm}
\includegraphics[width=\columnwidth]{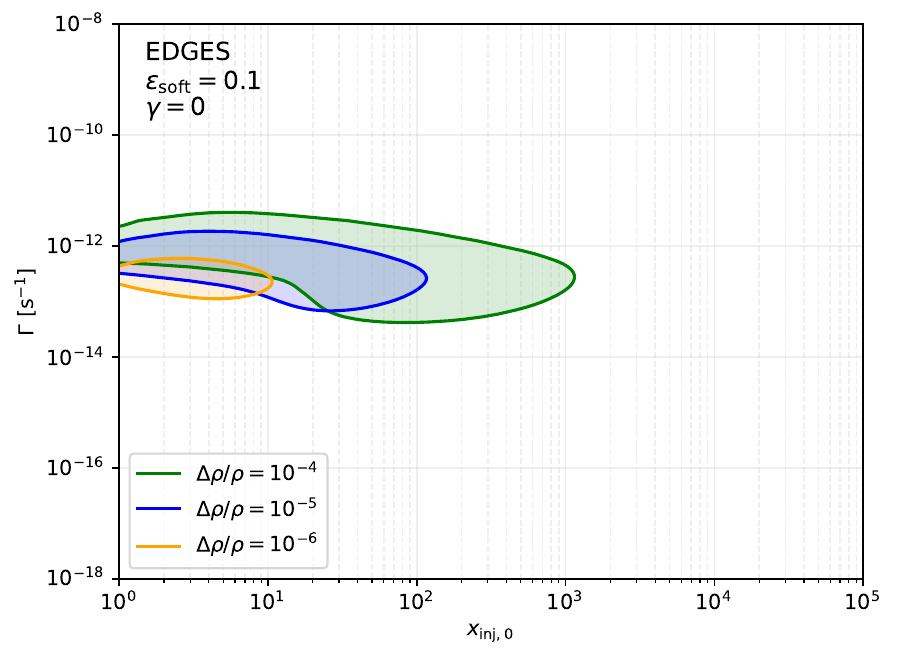}
\caption{Limits on $\Gamma$ and $x_{\rm inj,0}$ derived when taking the EDGES measurement as an upper bound on the global 21 cm absorption trough ($\Delta T_{\rm b}$. The presence of significant low-frequency photons produces a soft photon heating effect which makes the global 21 cm signal much more difficult to detect, thereby weakening constraints. The strip of constraints presented here correspond to the ``sweet spot" in which near recombination, low frequency photons are still absorbed by the background and do not induce this heating effect.}
\label{fig:EDGES_const}
\end{figure*}

\begin{figure*}
\centering 
\includegraphics[width=\columnwidth]{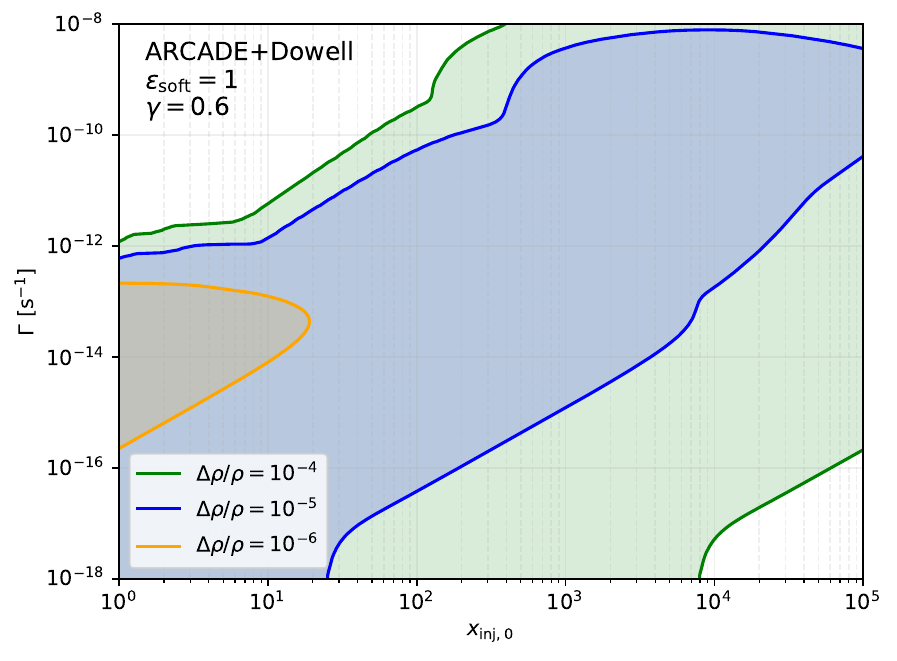}
\hspace{4mm}
\includegraphics[width=\columnwidth]{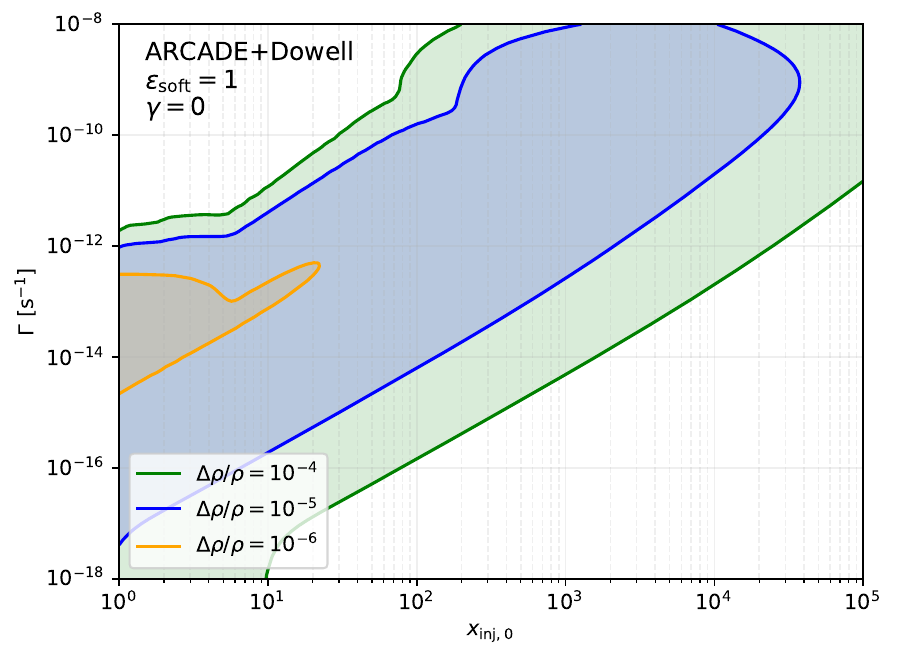}
\\[10mm]
\includegraphics[width=\columnwidth]{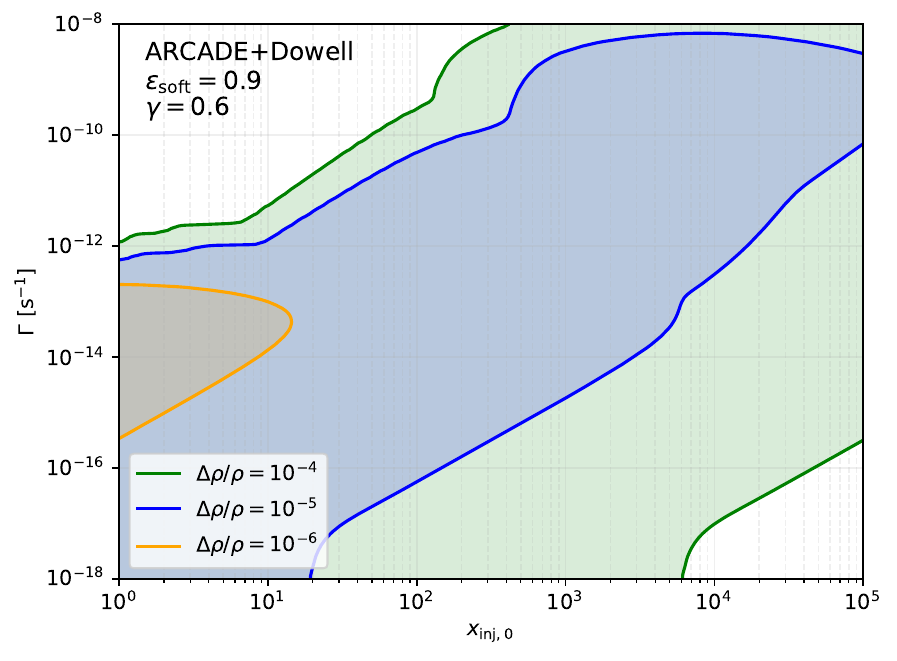}
\hspace{4mm}
\includegraphics[width=\columnwidth]{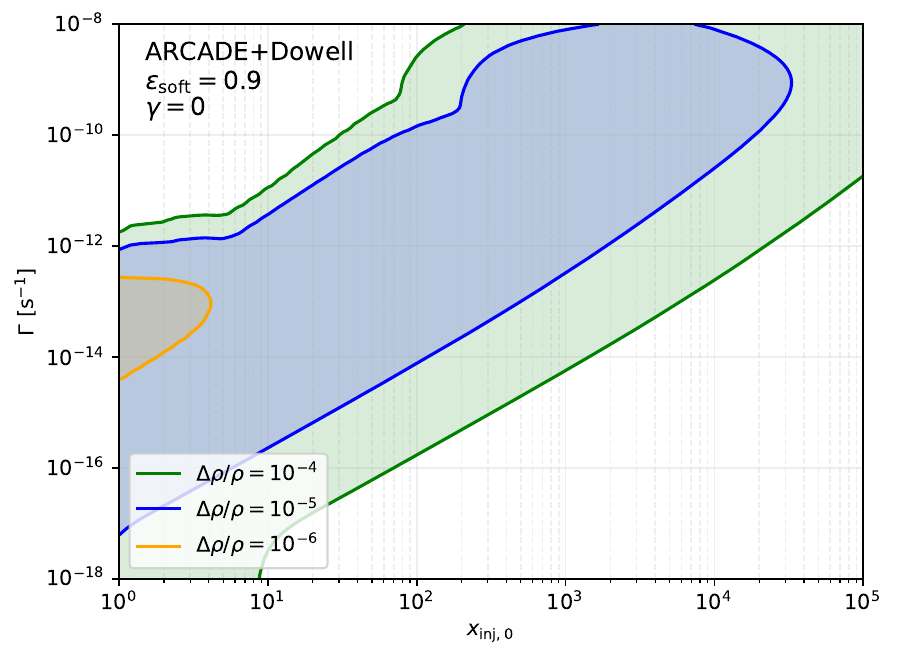}
\\[10mm]
\includegraphics[width=\columnwidth]{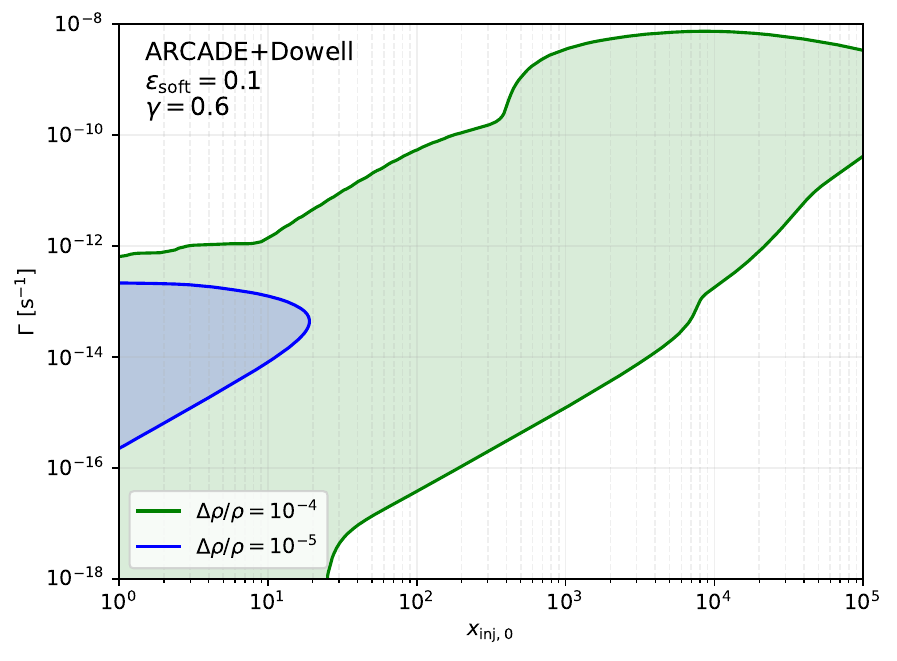}
\hspace{4mm}
\includegraphics[width=\columnwidth]{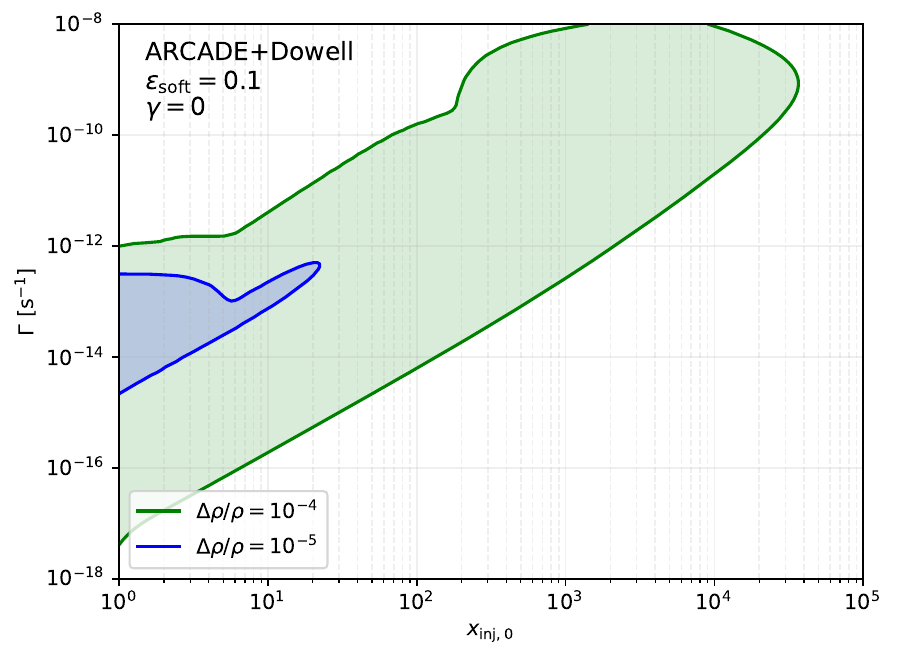}
\caption{Constraints from the production of a radio background in excess of the fit from \citet{Fixsen2011, DT2018} presented in Eq.~\eqref{eq:RSB-fit}}
\label{fig:AD_const}
\end{figure*}

In Fig.~\ref{fig:total_const}, we plot the exclusion region for both synchrotron and free-free injection spectral, taking $\epsilon_{\rm soft}=1$, $0.9$ and $0.1$. We combine constraints derived from CMB spectral distortions, CMB anisotropies, global 21 cm (the EDGES limit), and radio synchrotron background limits in this figure. We breakdown the individual constraints in Figs. \ref{fig:FIRAS_const}, \ref{fig:XeTau_const}, \ref{fig:EDGES_const} and \ref{fig:AD_const}. From these figures, it is clear that {\it COBE/FIRAS}, the RSB, and CMB anisotropy data constrain the parameter space in highly complementary ways. We describe these constraints in some detail below.

For $\Delta\rho_{\gamma}/\rho_{\gamma}=10^{-4}$, lifetimes shorter than $10^{13}$~s are strongly constrained by {\it COBE/FIRAS} as one would expect. For lower $\Delta\rho_{\gamma}/\rho_{\gamma}$, the constraints weaken significantly. However, we do see that shorter lifetimes and high $x_{\rm inj,0}$ are still constrained by spectral distortion data. Naively, one may expect that this should not be possible as {\it COBE/FIRAS} allows $y$ and $\mu$-distortions at the level of $10^{-5}$. However, as we have already mentioned, we do not assume the spectral distortion shapes to be just $y$ or $\mu$-type. With \texttt{CosmoTherm} it is possible to use the full spectral data from \COBEF to produce more general constraints. Spectral distortion constraints differ between synchrotron and free-free type injection, especially in post-recombination universe ($\Gamma\lesssim 10^{-13}$s$^{-1}$). During this epoch, thermalization is highly inefficient and one basically observes only a simple redshifting of the injected photons. The free-free injection spectrum is flatter, and therefore there are more photons visible in the {\it COBE/FIRAS} band compared to the synchrotron case. As we reduce $\epsilon_{\rm soft}$, the constraints starts to look similar for both cases. This is expected as we are simply reducing the branching ratio of energy into the soft photons. Pure energy (heat) injections do not contain direct spectral information and are thus much less stringently constrained by \COBEF for injections in the post-recombination epoch. Interestingly, we see a small strip in all three cases of $\epsilon_{\rm soft}$ for which the constraint is relaxed. As the non-thermal photons thermalize, more $y$-distortions are created which contain a characteristic negative feature at $\nu<217$ GHz (it is positive otherwise). In contrast, the directly injected photons always have positive intensity. This superposition leads to a cancellation which relaxes the constraints somewhat and gives rise to this interesting feature. This may also explain the weakening of constraints as we reduce $\epsilon_{\rm soft}$ from 0.9 to 0.1 for $\Delta\rho_{\gamma}/\rho_{\gamma}=10^{-5}.$

For $\Gamma\lesssim 10^{-13}$s$^{-1}$ ($z\approx 10^3$), the constraints become dominated by CMB anisotropy considerations (Fig. \ref{fig:XeTau_const}). In particular, we find severe constraints for $x_{\rm inj,0}\gtrsim 10^4$. This is due to the abundant production of significantly energetic photons which can directly ionize neutral hydrogen in the post-recombination era. The ionization threshold of $13.6$ eV corresponds to $x_{\rm inj,0}\approx 6\times 10^4$, but we remind the reader that we have a smooth exponential tail of higher energy photons in the injection spectrum, as can be seen in Eq.~\eqref{eq:soft_spectrum}. A small number of photons in the exponentially suppressed tail can yield reasonable constraints as the baryons are significantly outnumbered by the CMB photons. Therefore, even values of $\Delta\rho_{\gamma}/\rho_{\gamma} \simeq 10^{-8}$ can imply a sizable number of ionizing photons. For $x_{\rm inj,0}\lesssim 10^4$ and $\Delta\rho_{\gamma}/\rho_{\gamma}\lesssim 10^{-6}$, the constraints are generally weak as the number of ionizing photons drops substantially. However, the low frequency photons in these cases are absorbed by the electrons (soft photon heating) and heat the baryons. This changes the ionization history because the recombination coefficient is a sensitive function of gas temperature. Therefore, we do find some regions of parameter space ruled out for $x_{\rm inj,0}\lesssim 10^4$ by considering both CMB anisotropy and $\tau$ (reionization) constraints. This soft photon heating effect \citep{ACC2023} is more important for a synchrotron-like injection spectrum, as a steeper spectral index implies more soft photons. Therefore, we find stronger constraints for  the synchrotron case relative to a free-free spectrum. Again the $\epsilon_{\rm soft}<1$ scenarios look similar for both the cases and the constraints becomes stronger with decreasing value of $\epsilon$ (i.e. the addition of more heat). This is due to increases in the collisional ionization rates, as well as the reduction of the recombination coeffcient with the increase in gas temperature.

In Fig. \ref{fig:EDGES_const}, we plot the exclusion region determined from the EDGES limit which is restricted to a narrow strip around the recombination epoch ($\Gamma \simeq 10^{-13} \, {\rm s}^{-1}$). When injection happens before recombination, the low-frequency photons are rapidly absorbed by the highly charged background, leaving few photons to disturb $\Delta T_{\rm b}$.
%
%
For longer lifetimes, the injected radio background persists and can produce a large absorption signal. However, the heating due to soft photons raises the gas temperature, which ultimately reduces the amplitude of the absorption signal \citep{ACC2023}. Therefore, the constraints can again be relaxed. Soft photon heating is mediated by photons with $x\lesssim 10^{-4}$ which are too low frequency to directly produce a 21 cm radio background (i.e. they do not alter $T_{\rm R}$). 
There is a sweet spot around the time of recombination when photons with $x \lesssim 10^{-4}$ are still absorbed by the background, but photons around the cosmic dawn rest frame $21$ cm frequency freely propagate. For these cases alone, soft photon heating does not occur and the $21$ cm absorption signal is boosted, allowing us to place constraints from the EDGES limit.Readers are referred to \cite{ACC2023} for more details.


We show the constraint coming from extra radio backgrounds in Fig. \ref{fig:AD_const}. There seems to be a direct correlation between $\Gamma$ and $x_{\rm inj,0}$ for the exclusion region. High frequency photons injected at high redshifts show up today within the ARCADE+Dowell frequency bands which leads to tighter constraints. Synchrotron-type injections provide a somewhat tighter limit due to their steepness.


\section{Power-law injection histories}
\label{sec:plaw_constraints}

\begin{figure}
\centering 
\includegraphics[width=\columnwidth]{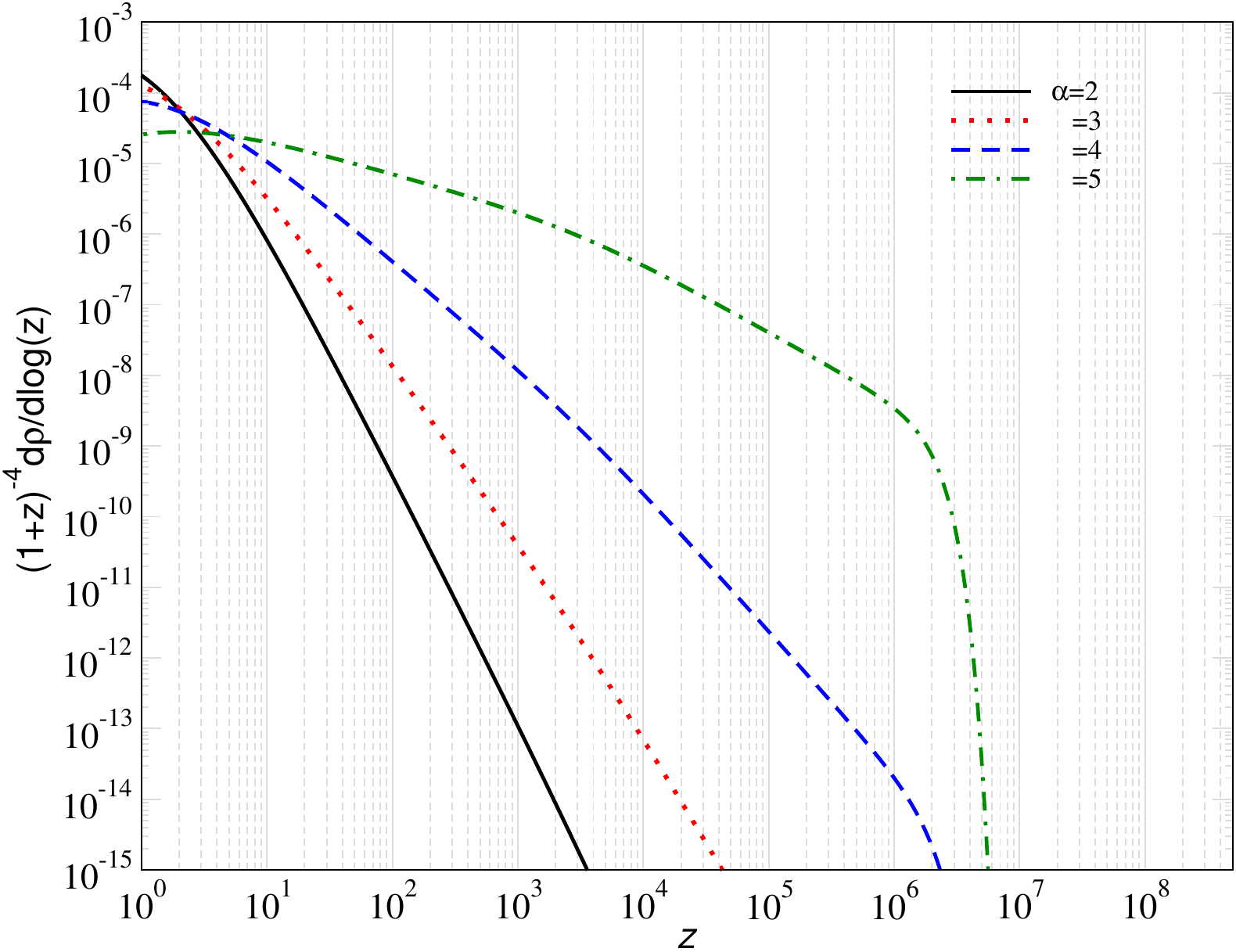}
\caption{A visualization of the injection history for power law injection cases with different values of the RSI ($\alpha$).}
\label{fig:plaw_injection}
\end{figure}

\begin{figure*}
\centering 
\includegraphics[width=\columnwidth]{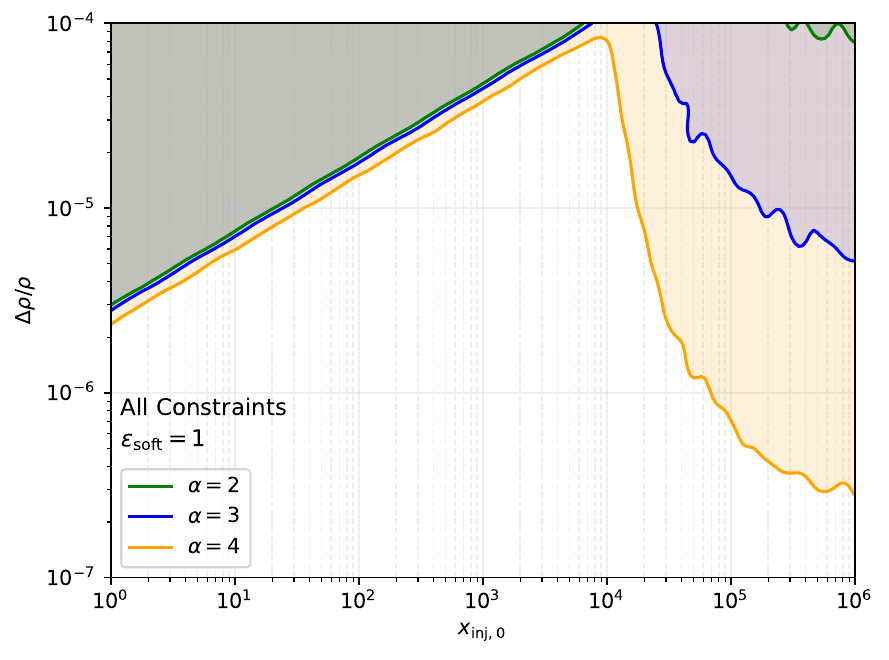}
\hspace{4mm}
\includegraphics[width=\columnwidth]{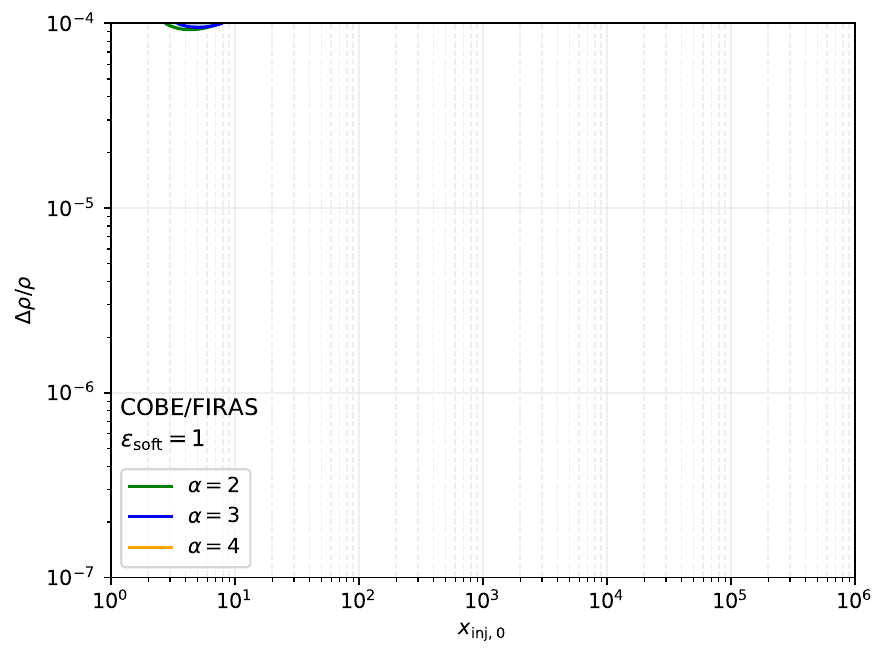}
\\[10mm]
\includegraphics[width=\columnwidth]{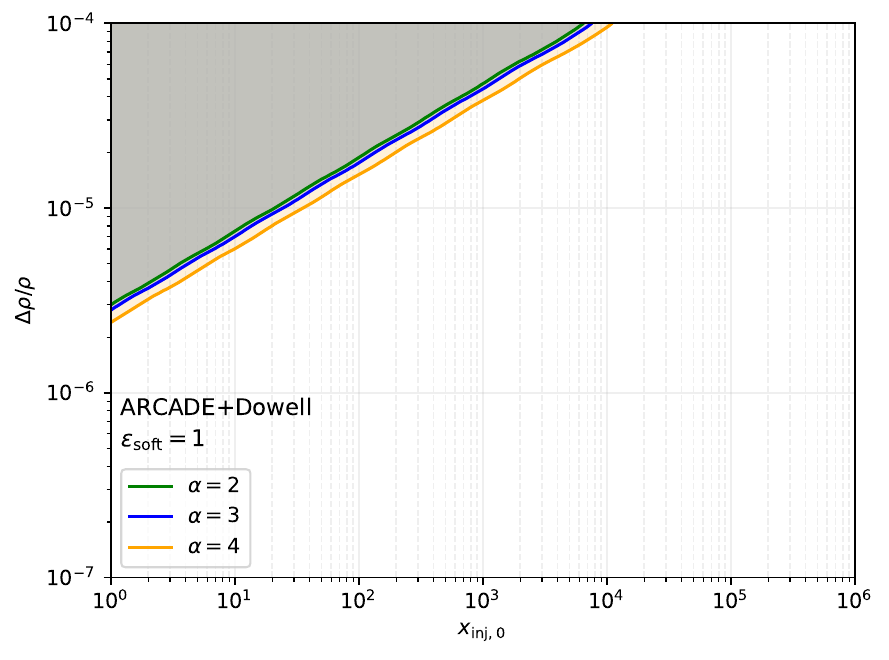}
\hspace{4mm}
\includegraphics[width=\columnwidth]{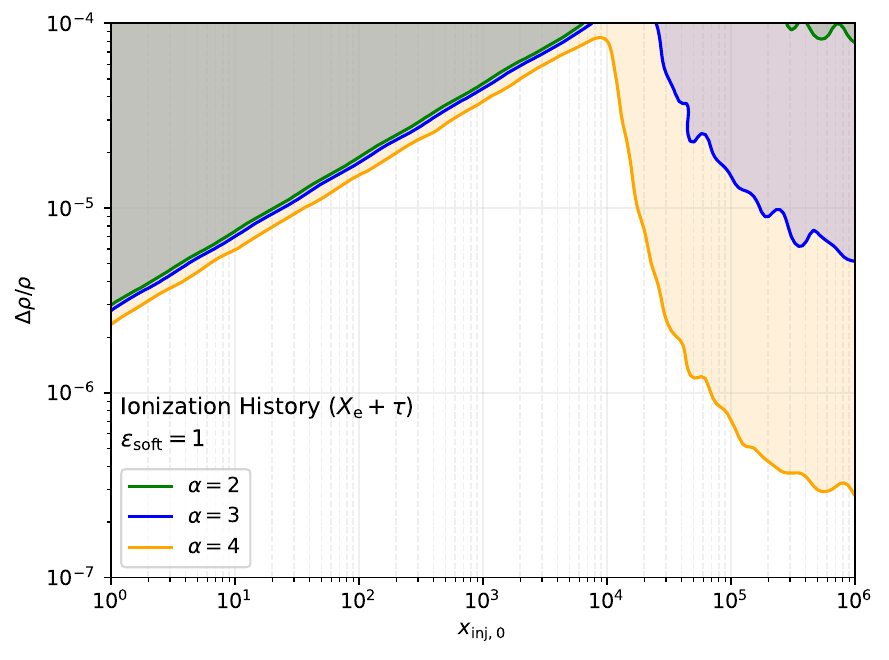}
\caption{Cumulative and partial constraints for power law injection scenarios with a synchrotron type spectrum and $\epsilon_{\rm soft}=1$ for different values of the RSI ($\alpha$).}
\label{fig:plaw_constraint}
\end{figure*}

In this section, we consider the case of a power-law injection in redshift, i.e.,
\begin{equation}
    \frac{{\rm d}\rho_{\gamma}}{{\rm d}t}\propto (1+z)^{\alpha} ,
    \label{eq:plaw}
\end{equation}
where the proportionality constant is determined by our choice of the total amount of energy injected ($\Delta\rho_{\gamma}/\rho_{\gamma}$). We refer to $\alpha$ as the redshift spectral index (RSI). The injected photon spectrum is kept the same as in previous case (that is, it is either synchrotron or free-free type). In Fig. \ref{fig:plaw_injection}, we show the rate of energy injection for a few different choices of the RSI. For $\alpha=3$, the case is similar to a decaying particle with lifetime longer than the universe. By changing this index, we probe either steeper or flatter energy injections (in redshift). For injections above $z\gtrsim 2\times 10^6$, the distortions are washed out, an effect that can be captured analytically using the distortion visibility function \citep{Hu1993, Chluba2011therm, Khatri2012b, Chluba2014}. 

In Fig. \ref{fig:plaw_constraint}, we plot the cumulative as well as the individual constraints. Similar to the decaying particle scenarios in Sec.~\ref{sec:decay_constraints}, we observe strong constraints for $x_{\rm inj,0}\gtrsim 10^4$. For larger values of $\alpha$, we have a flatter injection history (Fig.~\ref{fig:plaw_injection}) which affects the CMB anisotropies more significantly, allowing us to obtain stronger constraints. At $x_{\rm inj,0}\lesssim 10^4$, the constraints are driven by the RSB which puts a limit on the expected intensity of the signal. Since the intensity of the distortion is a degenerate combination of $\Delta\rho_{\gamma}/\rho_{\gamma}$ and $x_{\rm inj,0}$, the constraints satisfy a scaling relation.

We have kept the choice of $\alpha$ arbitrary in this section, making our discussion relatively model-independent, though we note that one can map many astrophysical scenarios to these test cases. For example, accreting black holes can emit radio photons with a synchrotron spectrum \citep{ECLDSM2018,MK2021}, which can be mapped to the $\alpha=3$ case since the number density of black holes redshifts like matter. 
For low-frequency photon emission, the radio synchrotron background provides the strongest constraints, and high redshift observables such as CMB spectral distortions and CMB anisotropy do not play an important role. However, if a correlated energetic UV photon emission exists, one expects strong CMB anisotropy constraints \citep{ADC2022}. We leave a more thorough investigation of this model to future work.

We can also consider astrophysical photon injection scenarios with a rate proportional to star formation rate, $\psi(z)\propto (1+z)^{-2.9}$ \citep{MD2014}. This is an extremely steep injection rate with $\alpha=-2.9$. Therefore, we do not expect significant constraints from CMB observables. We should remind the reader that all of these discussions apply for synchrotron-like and free-free type injection spectra. However, for different types of injection spectra (such as a $\delta$-function or a sharp injection profile), one needs to compute the constraints on a case-by-case basis \citep[e.g.,][]{Bolliet2020PI}. 

\section{Conclusions}
\label{sec:conclusion}
In this paper, we analyzed the evolution of broad photon spectrum injections over the history of the universe and obtained constraints on their parameter space from multiple cosmological probes. Compared to a more simple type of injection spectrum (e.g. $\delta$ injections), we expect more complicated physical processes to be at play. Using \texttt{CosmoTherm}, we have numerically followed these processes in order to compute a set of robust constraints on our model parameters.
%

As a simple case, we began by concentrating on a decaying particle scenario, but in the latter part of this work we also studied a more model independent power law injection histories in redshift. In the literature, the photon spectrum is typically assumed to be a $\delta$-function in order to simplify the calculations. Our work provides a necessary next step in going beyond this approximation. These calculations can be mapped to a wide variety of astrophysical and exotic phenomena, and capture the general property of having an injection of non-thermal photons which follow a power-law behaviour in both frequency and redshift.

\newpage
\vspace{-5mm}

\section*{Acknowledgments}

This work was supported by the ERC Consolidator Grant {\it CMBSPEC} (No.~725456).
JC was furthermore supported by the Royal Society as a Royal Society University Research Fellow at the University of Manchester, UK (No.~URF/R/191023).
BC would also like to acknowledge support from an NSERC-PDF.

\vspace{-5mm}
\section{Data availability}
The data underlying in this article are available in this article and can further be made available on request.

{\small
\vspace{-3mm}
\bibliographystyle{mn2e}
\bibliography{Lit}
}
\newpage



\end{document}